\documentclass[9pt,twocolumn,twoside]{opticajnl}

\journal{opticajournal}
\setboolean{shortarticle}{true}

\usepackage{comment}
\usepackage{outlines}
\usepackage{hyperref}
\hypersetup{
    colorlinks = true,
    linkcolor = blue,
    filecolor = magenta,
    urlcolor = cyan,
    pdftitle = {AIM Paper},
    pdfpagemode = FullScreen,
}
\usepackage{subfig}
\usepackage{caption}
\usepackage{setspace}
\usepackage{xcolor}
\usepackage{float}
\usepackage{soul}
\usepackage{listings}
\usepackage{lineno}

\usepackage{graphicx} 

\title{Implementing photonic-crystal resonator frequency combs in a photonics foundry}

\author[1,2,*]{Haixin Liu}
\author[1,2]{Ivan Dickson}
\author[3]{Alin Antohe}
\author[3]{Lewis G. Carpenter}
\author[1,2]{Jizhao Zang}
\author[1,2]{Alexa R. Carollo}
\author[1,2]{Atasi Dan}
\author[1]{Jennifer A. Black}
\author[1,2]{Scott B. Papp}

\affil[1]{Time and Frequency Division, National Institute of Standards and Technology, 325 Broadway, Boulder, Colorado 80305, USA}
\affil[2]{Department of Physics, University of Colorado, Boulder, Colorado 80309, USA}
\affil[3]{AIM Photonics, Albany, New York 12203 USA}
\affil[*]{Corresponding author: Haixin.Liu@colorado.edu}

\date{July 2024}

\begin{abstract}

We explore an AIM Photonics silicon-nitride platform to fabricate photonic-crystal resonators for generating optical parametric oscillators (OPO) and soliton microcombs. Our approach leverages the scalability and fine feature size of silicon-nitride processing on large-scale silicon wafers to achieve low-loss, high-Q microresonators, functionalized by nano-scale photonic-crystal structures. We demonstrate intrinsic microresonator quality factor up to $1.2\times10^7$ with complete foundry fabrication on 300 mm silicon, a 700 nm thick silicon-nitride device layer, and inclusion of complex nanophotonics. These features enable a host of nonlinear nanophotonics sources on the platform, including OPOs, microcombs, parametric amplifiers, squeezed-light generators, and single-photon sources. By fine-tuning the photonic-crystal design parameters, we achieve broad tunability in the frequency of the OPO output, spanning a significant portion of the near-infrared. Additionally, we observe the formation of soliton frequency combs, enabled by the precise dispersion engineering of the microresonators. These results highlight the potential of widely accessible, photolithographically patterned, silicon-nitride photonics to enable wide access to and complex integration of frequency-comb sources, with applications in spectroscopy, metrology, and communications.
\end{abstract}

\setboolean{displaycopyright}{true}

\begin{document}
\maketitle

\noindent Kerr-nonlinear microresonators provide frequency conversion of a laser from one frequency to others, which can happen in the form of optical parametric oscillators (OPOs) or soliton microresonator frequency combs (microcombs) \cite{tobias}. The signal and idler frequencies of an OPO provide coherent sources with by-design access to arbitrary wavelength, which is necessary in atomic and molecular spectroscopy and atomic clocks  \cite{lu2019efficient}. Soliton microcombs have myriad applications like optical frequency division \cite{Sun_2024}, optical-frequency synthesis \cite{spencer2018optical}, and optical data communication \cite{yang2022multi}. However, the frequencies of signal and idler waves of OPOs as well as the occurrence of solitons are subject to a narrow range of phase-matching, which traditionally can only be realized in microresonators with anomalous group-velocity dispersion (GVD) and requires high accuracy in the design and fabrication of the waveguide geometries \cite{lu2020chip}. In the past few years, photonic-crystal resonators (PhCRs) have emerged as a versatile platform for manipulating light at the nanoscale. 

A photonic crystal (PhC) is a dielectric structure of which the refractive index has a subwavelength spatial periodicity. In a microresonator, this can be in the form of a sinusoidal (or other shape \cite{Lu:24}) modulation of the ring width (RW), creating a varying effective refractive index \cite{Su-peng2021}. The two critical parameters of a PhC are its repetition $m_\text{PhC}$ and the peak-to-valley amplitude (APhC). For an even $m_\text{PhC}$, the PhC induces an interaction between clockwise (CW) and counterclockwise (CCW) propagation of azimuthal mode number $m=m_\text{PhC}/2$ inside the resonator, spkitting the frequency of this mode an amount $\Gamma$ determined by APhC \cite{Su-peng2021,black2022optical}. This technique enables the precise control of phase-matching in both anomalous and normal GVD, namely, universal phase-matching \cite{liu2024threshold}. So far, OPOs with span over an octave and output power of tens of mW have already been demonstrated in PhCR with near-zero and normal GVD \cite{brodnik2024,liu2024threshold}. It has also been found that PhCRs enable spontaneous soliton formation in anomalous GVD and stable solitons with high-efficiency and flat-top spectra with controllable bandwidth \cite{Su-peng2021,zang2024laserpower,spektor2024photonic}.

As PhCRs provide a robust and flexible mechanism for frequency conversion, it is useful to fabricate them in a repeatable and scalable way. Nowadays, commercial foundries like Ligentec, AIM Photonics, and Tower Semiconductor provide fabrication of photonic integrated circuits with high volume and consistent tolerance. A signature, long-term goal of foundry fabrication for optical-frequency metrology is a single-chip, self-referenced microcomb system, which would require an octave-spanning soliton plus f-2f detection and down-conversion of millimeter-wave repetition-rate signals \cite{zang2024foundry}. High-performance data-communication links and photonic AI computing accelerators are another clear driver to develop foundry accessible microcomb sources \cite{Feldmann_2021}. PhCRs can enhance microcomb performance in these applications and others \cite{Su-peng2021,zang2024laserpower,10210009}, but the potential of foundry manufacturing PhCRs has not been comprehensively investigated. Implementing PhCRs requires definition of sub-wavelength structures of periodicity $\Lambda=2\pi R/m_\text{PhC}$, where $R$ is the microresonator radius, with nanometer scale modulations of the RW, while maintaining high quality factor and more critically high external coupling rate. Hence, PhCR fabrication leverages the capabilities of foundry processing to simultaneously create ultra-smooth and low loss resonators in the over-coupled regime, functionalized by moderate-resolution nanostructures with extremely precise size. 


Here, we present silicon-nitride integrated-photonics fabrication of high-Q (up to $1.2\times10^7$) PhCRs on a large-scale silicon wafer at AIM Photonics, which offers numerous capabilities highlighted by the silicon-photonics foundry. Through precise control of RW, $m_\text{PhC}$, and APhC, we systematically engineer the GVD and PhC mode split and achieve widely controllable OPO and soliton microcombs in both the anomalous and normal GVD regimes. This demonstrates comprehensive access to nonlinear phase matching and fabrication in a widely available foundry.  We characterize the relation between the OPO span and $\Gamma$, which agrees well with our designs, demonstrating precise control of phase-matching in PhCRs. Our work highlights the ability of a photonics foundry to extend the potential of PhCRs in advancing compact, controllable laser sources and broadening the applications of frequency combs in various fields.

\begin{figure}[!htb]
    \centering
    \includegraphics[width=\linewidth,trim={0cm 0cm 0cm 0cm}, clip]{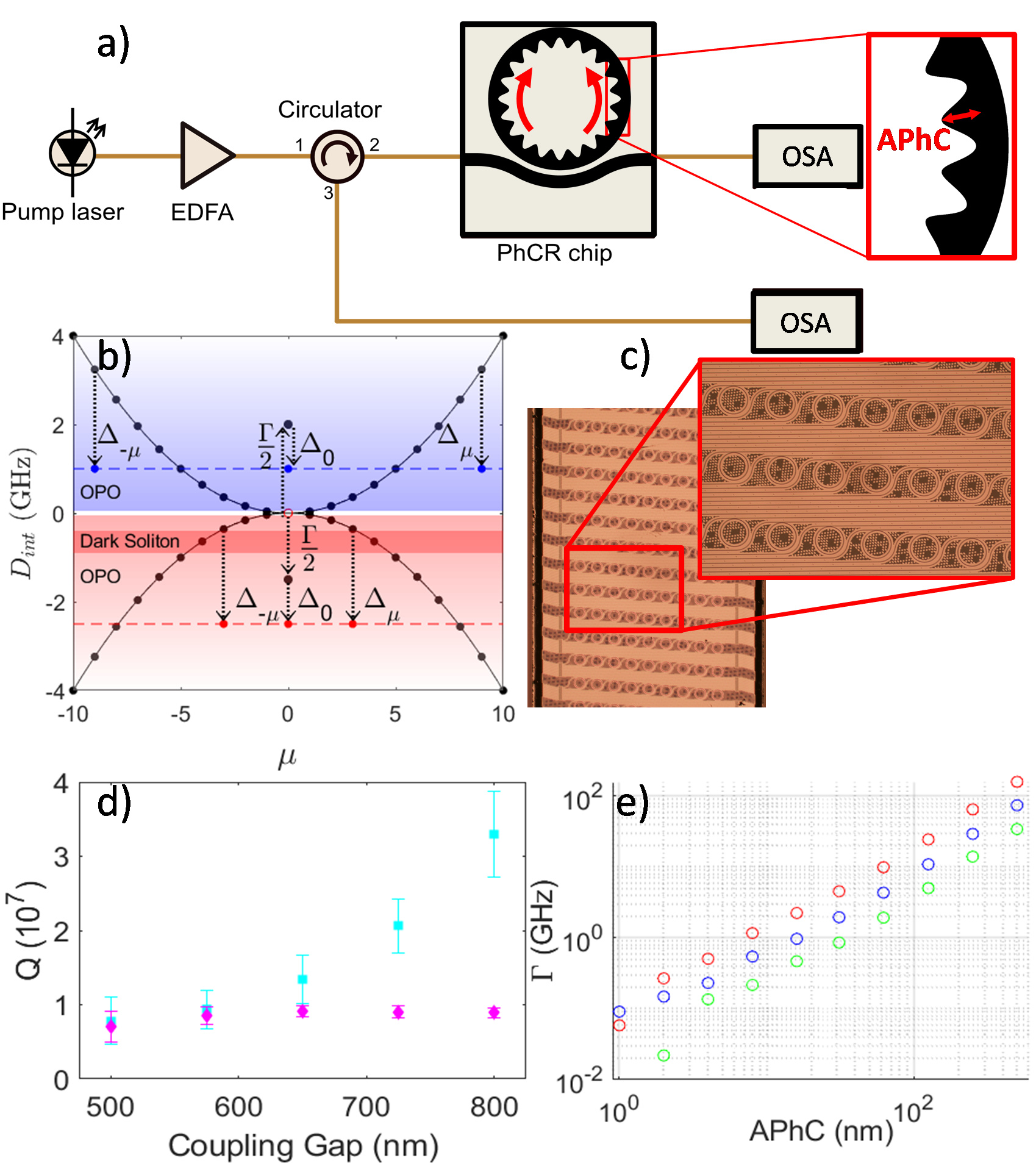}
    \caption{(a) Chip testing setup. The inset illustrates our PhC structure. A circulator collects the light from both directions. (b) The phase-matching diagram of PhCRs. $D_{int}$ for an arbitrary anomalous GVD (blue) and an arbitrary normal GVD (red) are plotted. The mode split is denoted by $\Gamma$ and the Kerr shift of mode $\mu$ is denoted by $\Delta_\mu$. Two dashed lines show the phase-matching condition. The red region shows the range a dark soliton exists. (c) Microscope image of chip and microresonator device. (d) Measured $Q_{i}$ (magenta) and $Q_{c}$ (cyan) for various coupling gaps during the wavelength scan from 1580 nm to 1615 nm. (e) $\Gamma$ v.s. APhC for devices with RWs of 1.7$\mu$m (red), 2.0$\mu$m (blue), and 2.3$\mu$m (green).}
    \label{Fig:1}
\end{figure}
To implement PhCR OPOs and microcombs, we design devices for a 700 nm thick, custom stoichiometric silicon nitride (SiN) layer and submit them to AIM Photonics for fabrication. We base our designs on the standard reference value for SiN index of refraction \cite{Luke:15} and previous experiments with fully silicon dioxide clad SiN microresonators \cite{PhysRevApplied.11.044017,zang2024foundry}. To comprehensively demonstrate the usability of foundry manufactured PhCRs, our design includes different RWs: 1.7 $\mu$m, 2.0 $\mu$m and 2.3 $\mu$m, which enables access to both anomalous and normal GVD. Moreover, we design the resonator coupling gap in accord with FDTD simulations. We implement PhCRs through a periodic inner sidewall grating of the RW, using APhC to control the coherent backscattering rate in the resonator and hence $\Gamma$ of mode $m=m_\text{PhC}/2$.

For optical testing of PhCRs, 
we measure the GVD and quality factor as well as the properties of the OPO and soliton microcombs generated in the devices, and we compare the results with our design and simulation targets. Figure 1(a) is a schematic diagram of our experimental setup for device characterization and OPO or comb generation. The pump is generated by a widely tunable continuous-wave laser at 1550 nm, which is followed by an erbium-doped fiber amplifier (EDFA) to boost the optical power. Since the field inside PhCR combs are bidirectional, we use an optical circulator to separate the forward and the backward propagating light. The pump is coupled into and out of the PhCR chip through lensed fibers with a coupling loss of 2 dB per side. Two optical spectrum analyzers (OSAs) monitor the forward and backward propagating pulses generated by the PhCR. The inset of Fig. \ref{Fig:1}(a) is an enlarged view of the PhCR, emphasizing the periodical modulation of the ring width.

Figure 1(b) illustrates the operating principle of nonlinear phase matching in a PhCR with an anomalous GVD case shown in the upper portion of the figure and a normal GVD case shown in the lower portion. The GVD here is characterized by the integrated dispersion, $D_\text{int} = \nu_\mu - (\nu_0 +\text{FSR}\mu)$, where FSR is the free spectral range, $\mu$ is the mode number relative to the pump and $\nu_\mu$ is the cold resonance frequency of the PhCR \cite{black2022optical}. The split mode $m$ coincides with the pump, and we indicate $\Gamma$ and the Kerr shift of each mode, represented by $\Delta_\mu$ \cite{Su-peng2021}. 
We achieve phase-matching by pumping the blue shifted mode in the anomalous GVD rings and the red shifted mode in the normal GVD rings, which are shown by the blue and red dashed traces, respectively. For anomalous GVD, we scan $\Gamma$ so that the pump mode is phase-matched to different modes, enabling a controllable signal-to-idler frequency span ($\Delta\nu_{\text{OPO}}$). For normal GVD, while the devices with large $\Gamma$ behave similarly to the anomalous GVD case, for small $\Gamma$ (shaded red region in Fig. \ref{Fig:1}(b)), both our simulation and experiment show dark soliton formation following OPO as we increase the detuning. 


The AIM Photonics foundry fabricates our devices at the target SiN device layer dimensions with $\sim$5 $\mu$m of silicon oxide lower cladding and a full upper cladding; see Fig. 1(c) for an image of a chip with a high packing density of PhCRs. The fabrication process is typical for SiN microresonators, including LPCVD to create the SiN device layer, high temperature annealing to reduce SiN absorption, and top oxide cladding deposition. To reduce optical absorption from the top silicon oxide cladding, we performed a post-process anneal at 500$^\circ$C in air for twelve hours.  High resolution stepper lithography allows for small features, crucial for writing the PhCRs with a $\Lambda\sim$ 400 nm period and a high density of microresonators to vary APhC ranging from 1-500 nm. All the microresonators have $R$ of 45 $\mu$m, which corresponds to FSR $\sim$500 GHz. Some hydrogen absorption is observed in the fabricated resonators, reducing the Q of resonances at $\lambda$ less than 1580 nm \cite{Ye:23,Liu_2021,10.1063/5.0057881}. Therefore, the Q factor measurements we report are obtained from data in the 1580 nm to 1615 nm bands. Future optimization of thermal annealing in the silicon nitride foundry process can likely reduce hydrogen content and improve the Q-factor across the telecom band.

A sufficiently high intrinsic quality factor $Q_\text{i}$ is important in nonlinear microresonators to set the threshold power at a manageable level and to facilitate higher conversion efficiency of both OPO and soliton microcombs. 
While the coupling quality factor $Q_\text{c}$ depends the coupling rate, $Q_\text{i}$ characterizes the actual power dissipation of microresonators. For chip-based microresonators, SiN waveguides with silica cladding are widely adopted due to its low intrinsic material loss and integration compatibility, providing a potential high $Q_\text{i}$ \cite{Wu:20}. 
In our experiment, $Q_\text{i}$ is measured by characterizing the transmission spectrum measured by a photodiode at a pump power much lower than the threshold for OPO. A polarization controller is inserted before the chip and we specify our measurement on the fundamental TE mode family. We scan the pump wavelength from 1580 nm to 1615 nm, and for each resonance dip, the resonance frequency $\nu$, the full-width-at-half-maximum linewidth $\delta\nu$, and the minimum transmission $T$ are recorded. Then, $Q_\text{i}$ and $Q_\text{c}$ are solved using the following equation \cite{Wu:20,black2022optical}:
\begin{equation}
\begin{split}
&\frac{1}{Q_\text{i}}+\frac{1}{Q_\text{c}}=\frac{\delta\nu}{\nu}\\
&T=\left(\frac{Q_\text{c}-Q_\text{i}}{Q_\text{c}+Q_\text{i}}\right)^2
\end{split}
\end{equation}
Notice that $Q_\text{i}$ and $Q_\text{c}$ are symmetrical in the above equation. To tell them apart, we test the devices with different gaps. The $Q_\text{c}$ is expected to increase exponentially over the gap, while $Q_\text{i}$ should have much smaller variation.
Figure \ref{Fig:1}(d) shows the average values of $Q_\text{i}$ (magenta diamonds) and $Q_\text{c}$ (cyan squares) of forty PhCRs. These devices have five different gap values (500 -- 800 nm), eight APhC values (1 -- 125 nm) and RW = 2.3 $\mu$m. The error bar of each data point characterizes the standard deviation of quality factors during the wavelength scan and APhC scan. As expected, $Q_\text{c}$ increases a lot when the gap increases and $Q_\text{i}$ varies much less. Significantly, the highest $Q_\text{i}$ measured in our experiment is $(12\pm1)\times10^6$ with gap = 725 nm and APhC = 31 nm, which is much higher than one previously reported value ($Q_\text{i}=1.4\times10^6$) of another commercial silicon-nitride foundry \cite{zang2024foundry}. This indicates that a commercial foundry can achieve a high $Q_\text{i}$ PhCR despite the fine sidewall modulation.

Deterministic control of $\Gamma$ is critical to perform precise phase-matching. In order to precisely control $\Gamma$ through an APhC scan, a monotonic relation between them is necessary. Recent research in air-clad silicon-nitride PhCR reports a non-monotonic increase of $\Gamma$ over APhC with a bandgap closing point at APhC $\sim10^2$ nm, due to the degeneracy of the wider effectively and the narrower effectively modes \cite{Lu:24}. Here, we investigate the same $\Gamma$-APhC relation in our silicon-dioxide-clad devices and characterize $\Gamma$ of these PhCRs for the phase-matching research later. By scanning the pump frequency over the split-mode resonance, $\Gamma$ is measured as the spectral distance between the two transmission dips. Figure \ref{Fig:1} (e) plots the measured $\Gamma$ over a varying APhC of 1-500 nm with three different RWs of 1.7$\mu$m (red), 2.0$\mu$m (blue) and 2.3$\mu$m (green). The result shows a monotonic increase of $\Gamma$ over the whole APhC scan range and a power-law dependence of $\Gamma$ on APhC with deviation at very small (several nm) APhC for all three RW's. The tuning range of $\Gamma$ is from 0.1 GHz to more than 100 GHz, which enables applications of PhCRs from high efficiency dark soliton generation to broadband tunable OPO \cite{zang2024laserpower,brodnik2024}. For a larger RW, APhC needs to be increased correspondingly to have the same $\Gamma$. Our result demonstrates that the silicon dioxide cladding effectively improves the monotonic relation between $\Gamma$ and APhC of silicon-nitride PhCR. 

\begin{figure}[!htb]
    \centering
    \includegraphics[width=\linewidth,trim={0cm 0cm 0cm 0cm}, clip]{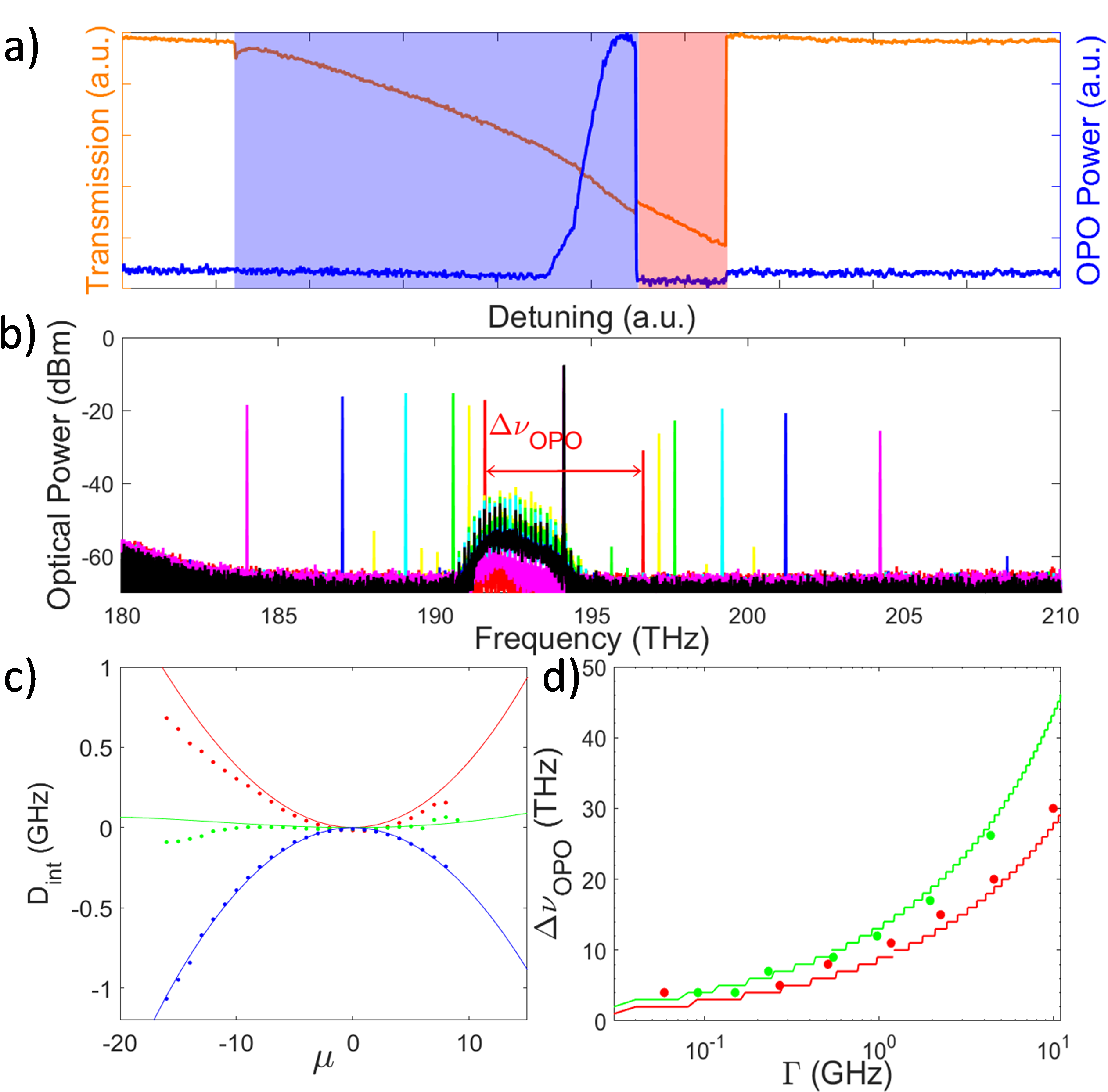}
    \caption{OPO in anomalous GVD PhCR. (a) The transmission of pump laser (orange curve) and OPO (blue curve) in the forward direction over detuning sweep. The blue and red shaded regions are the blue and red detuned resonances. (b) OPO spectra measured off-chip through component coupling losses of RW = 1.7 $\mu$m devices with six $\Gamma$'s ranging from 50 MHz (red) to 4.5 GHz (magenta). (c) Simulated $D_{int}$ (curves) and measured $D_{int}$ data (dots) for RW = 1.7 $\mu$m (red), 2.0 $\mu$m (blue), and 2.3 $\mu$m (green). (d) The measured $\Delta\nu_{OPO}$ (dots) over $\Gamma$ for RW = 1.7 $\mu$m (red) and 2.0 $\mu$m (green) compared to simulation (curves).}
    \label{Fig:2}
\end{figure}
We now demonstrate the precise control of phase-matching in foundry-manufactured PhCRs. We first investigate the anomalous GVD case, increasing the pump power above threshold and sweeping the pump frequency from the blue-shifted mode (negative detuning) to positive detuning. Figure \ref{Fig:2} (a) illustrates the measured transmission in the forward direction over detuning of the pump power (orange) and the OPO power (blue) with pump filtered. The OPO is generated in the thermal triangle of the blue-shifted mode (the blue-shaded region), and collapses when the detuning approaches the red-shifted resonance (the red-shaded region), which is consistent with the phase-matching diagram in Fig. \ref{Fig:1} (b). Next, we manually tuned the laser frequency into the resonance and recorded the onset of OPO sidebands at threshold pump power with an OSA. Figure \ref{Fig:2} (b) shows the OPO spectra of devices with RW = 1.7 $\mu$m and six $\Gamma$'s varying from 50 MHz (red) to 4.5 GHz (magenta). For further investigation of the $\Delta\nu_{\text{OPO}}$ - $\Gamma$ relation, we measured the GVD of rings with all three RWs by sweeping the pump wavelength from 1515 nm to 1615 nm and recorded the cold cavity resonance frequencies. Figure \ref{Fig:2} (c) shows the results and the simulated $D_\text{int}$ using COMSOL with RW = 1.7 $\mu$m (red), 2.0 $\mu$m (green), and 2.3 $\mu$m (blue). The dots with the corresponding colors are the measured $D_\text{int}$ with the same RWs, showing that the devices fabricated by the commercial foundry achieve an agreement with our design in GVD. We further measured $\Delta\nu_{\text{OPO}}$ at threshold on devices with different RWs and $\Gamma$. Figure \ref{Fig:2} (d) plots the dependence of $\Delta\nu_{\text{OPO}}$ on $\Gamma$. The colored curves are the predicted $\Delta\nu_{\text{OPO}}$ in the numerical simulation based on the modified Lugiato-Lefever equation \cite{liu2024threshold}. The dots with the corresponding colors are the measured $\Delta\nu_{\text{OPO}}$. We experimentally achieve $\Delta\nu_{\text{OPO}}$ of 2.5 THz to 30 THz with the $\Gamma$ varying from 50 MHz to 10 GHz. The good agreement between the predicted and measured $\Delta\nu_{\text{OPO}}$ demonstrate precise control of phase-matching in foundry-manufactured PhCRs through $\Gamma$. In principal, $\Delta\nu_{\text{OPO}}$ beyond an octave can be achieved with a larger $\Gamma$ \cite{brodnik2024}.

\begin{figure}[!htb]
    \centering
    \includegraphics[width=\linewidth,trim={0cm 0cm 0cm 0cm}, clip]{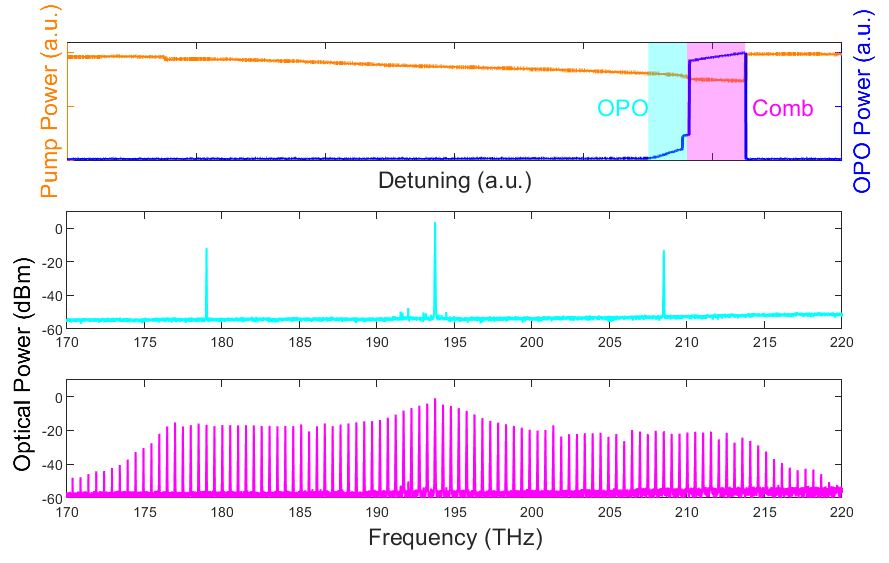}
    \caption{OPO and soliton microcomb generated by tuning the pump laser across the red-shifted split mode of a normal GVD device with $\Gamma= 860$ MHz and measured off-chip. (a) Pump-laser detuning sweep, showing the pump power and the power of newly generated light. The sweep starts between the red and blue mode, hence we independently excite the red-shifted mode. The shaded regions indicate OPO and soliton microcomb generation, respectively. (b) OPO spectrum at lower pump detuning (c) Frequency comb spectrum at larger pump detuning.}
    \label{Fig:3}
\end{figure}
In the normal GVD case (RW = 2.3 $\mu$m), the PhCR not only enables OPO with controllable frequency span, but also allows for stable soliton generation. To investigate this behavior, we tune the pump wavelength to the red-shifted resonance, and then increase the detuning. Different from the anomalous GVD case, most of the OPO (soliton) power comes out in the backward direction. The upper panel of Fig. \ref{Fig:3} shows the output power of the pump frequency in the forward direction and OPO (soliton) in the backward direction over detuning of a normal GVD PhCR with RW=2.3 $\mu$m and $\Gamma$ = 960 MHz. As detuning increases, we first observe OPO (the cyan shaded region) with $\Delta\nu_\text{OPO}$ = 30 THz. The middle panel of Fig. \ref{Fig:3} is its spectrum. Further increase of detuning leads to spontaneous formation of a backward-propagating soliton, highlighted by the magenta shaded region in Fig. \ref{Fig:3} upper panel. While there is an unsteady chaos stage prior to the soliton formation in ordinary resonators with anomalous GVD, which requires extra techniques such as fast sweeping to access the stable soliton regiem \cite{zang2024foundry}, in a PhCR with normal GVD the soliton is formed stably through the cascade of OPO, thus simplifying the operation. Moreover, compared with the traditional soliton in anomalous GVD, the soliton spectrum in normal GVD PhCR (see Fig. \ref{Fig:3} lower panel) features a flatter spectral power profile, spanning from 177 THz to 212.5 THz in this device. Previous research has demonstrated that the spectral width of the soliton can be controlled by $\Gamma$ \cite{spektor2024photonic}. 

In summary, we utilize the scalability and fine feature size offered via by a silicon-nitride platform under development in a silicon integrated-photonic foundry to fabricate microresonators with embedded nano-scale photonic crystals and a quality factors up to $1.2\times10^7$. By engineering GVD and $\Gamma$ through a systematic parameter scan of RW and APhC, we successfully realize precise control of phase-matching in both anomalous and normal GVD regimes, the generation of OPO with tunable frequencies, and the stable formation of soliton frequency combs. These results were obtained with one single tape-out, which focused on exploring a wide-range of $\Gamma$, and future work will target more specific comb states. Our experiments demonstrate the versatility and effectiveness of PhCRs and highlight the photonics foundry as a powerful toolkit for optical communications, spectroscopy, and optical metrology.  

\vspace{2.5mm}
\fontsize{8}{8}\selectfont\noindent \textbf{Funding.} Air Force Office of Scientific Research (FA9550-20-1-0004); National Institute of Standards and Technology; National Science Foundation (OMA – 2016244); and PIC fabrication via AIM Photonics (agreement number FA8650-21-2-1000).

\vspace{2.5mm}
\noindent \textbf{Acknowledgement.} We thank 
Grant Brodnik and Lindell Williams for carefully reviewing the manuscript, and Nick Usechak for helpful comments and providing us with chip space on AIM Photonics. This work is a contribution of the U.S. government and is not subject to copyright. Trade names provide information only and not an endorsement. The views and conclusions contained herein are those of the authors and should not be interpreted as necessarily representing the official policies or endorsements, either expressed or implied, of the United States Air Force, the Air Force Research Laboratory or the U.S. Government. 

\vspace{2.5mm}
\noindent \textbf{Disclosures.} The authors declare no conflicts of interest.

\vspace{2.5mm}
\noindent \textbf{Data availability.} Data underlying the results presented in this paper are not publicly available but may be obtained from the authors upon reasonable request.

\bibliography{sample,references}
\bibliographyfullrefs{sample,references}
\end{document}